\documentclass[smallextended]{svjour3}       
\smartqed  
\usepackage{graphicx}
\begin{document}

\title{Preliminary study of the missing mass spectra via the $^{12}C(p, K^0_s)$ and $^{12}C(p,\Lambda)$ reactions  at 10 GeV/c. 
}


\author{P.Zh.Aslanyan
}


\institute{JINR \at Joliot Curie 6, Moscow Region,141980 Dubna, Russia\\
              \email{paslanian@jinr.ru}
}

\date{Received: date / Accepted: date}

\maketitle

\begin{abstract}

The missing mass spectra for the $^{12}C(p, K^0_s)$ and $^{12}C(p,\Lambda)$
reactions have been studied by using of the propane bubble chamber(PBC) data from 700000 stereo
photographs or $10^6$ inelastic interactions. The momentum spectrum of $\pi^-$ in range of 100-200 MeV/c
have observed the significant enhancement from the p+C$\to \pi^-\Lambda$X(p+C$\to\pi^-K^0_s$X) reaction.
The missing mass spectra  have been observed signals  for  the $p(p,\pi^-K^0_s)$, $^3H(p,pK^0_s)$,
p(p,$\Lambda)\pi^-\gamma$  and p(p,$\pi^-)\Lambda$  reactions.
This experimental study will need to continue by a different method of identification for a reaction channels.

\keywords{hyperon \and hypernucleus \and strangeness \and 4$\pi$ geometry}
\PACS{13.75.-n1 \and 13.75.Ev   \and 13.75.Jz \and 14.20.Jn \and 25.80.Nv \and 25.80.Pw}
\end{abstract}

\section{Introduction}
\label{intro}

The properties of hypernuclei reflect the nature of the underlying baryon-baryon
interactions and, thus, can provide tests of models for the free-space hyperon-nucleon
(Y,N) and hyperon-hyperon (Y,Y) interactions\cite{gal}. The missing mass(MM) method makes
it possible to obtain unique information on the masses and the structure nuclei.
The possibility to produce hypernuclei in the (p,$K^+$) reaction was firstly mentioned by \cite{china}.In fact, the recent studies on the (p ,$K^+$) reaction confirm a quite substantial production of associated $\Lambda$-hyperons  leading to production cross sections for $\Lambda$-hypernuclei in
the order of a few 100 $\mu$b  for p + Pb at 1.5 - 1.9 GeV\cite{rudy}.
The paper presents the preliminary experimental results from the MM spectra in p+C interactions\cite{panda11}. This analysis with the MM is the first step to explore  hadronic systems with strangeness.The  event by event analysis will be the next step.

\section{Experimental data }
\label{sec:1}

The events with $V^0$ ($\Lambda$ and $K^0_s$)  were identified
by using the criteria\cite{pepan09}. The mass of the identified 9838-events with $\Lambda$ hyperon and
4964-events with $K^0_s$ mesons is consistent with their PDG values. The FRTIIOF model  and experimental data comparison shown that there are observed significant enhancement for $\Lambda$ hyperons production in ranges of the scattering $\theta <0$ and azimuth $\phi\approx$ 0 angles \cite{panda11}
in the spherical system of coordinates. The  missing mass  error  is equal to $\approx$ 80-100 MeV/$c^2$ for the p(p,$K^0_s$) and p(p,$\Lambda$) reactions.

\section{The  missing mass spectra with $K^0_s$ meson}
\label{sec:2}

\begin{figure}
    \includegraphics[width=0.25\textwidth,height=0.15\textheight]{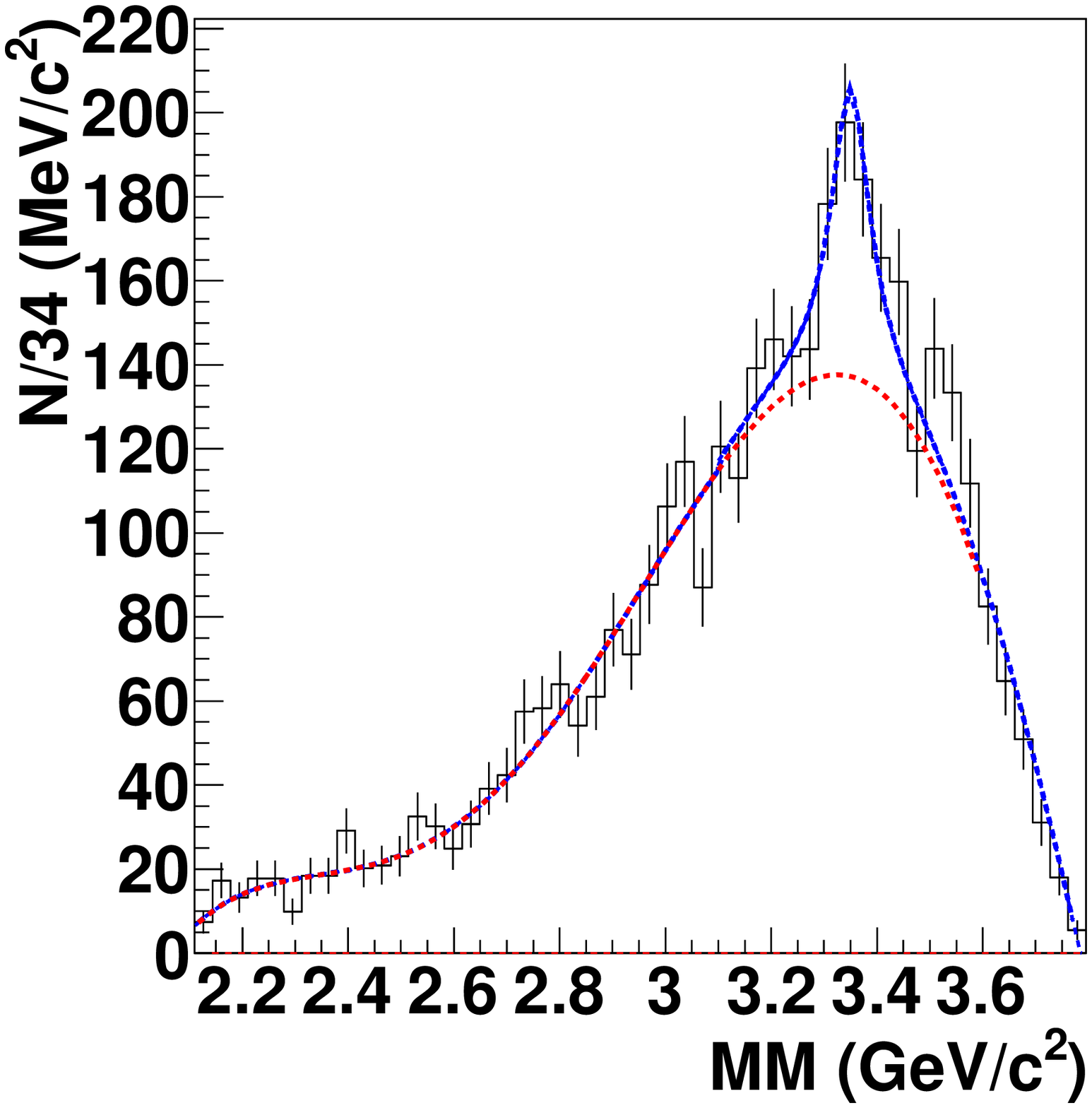}
    {\large a}
    \includegraphics[width=0.325\textwidth,height=0.15\textheight]{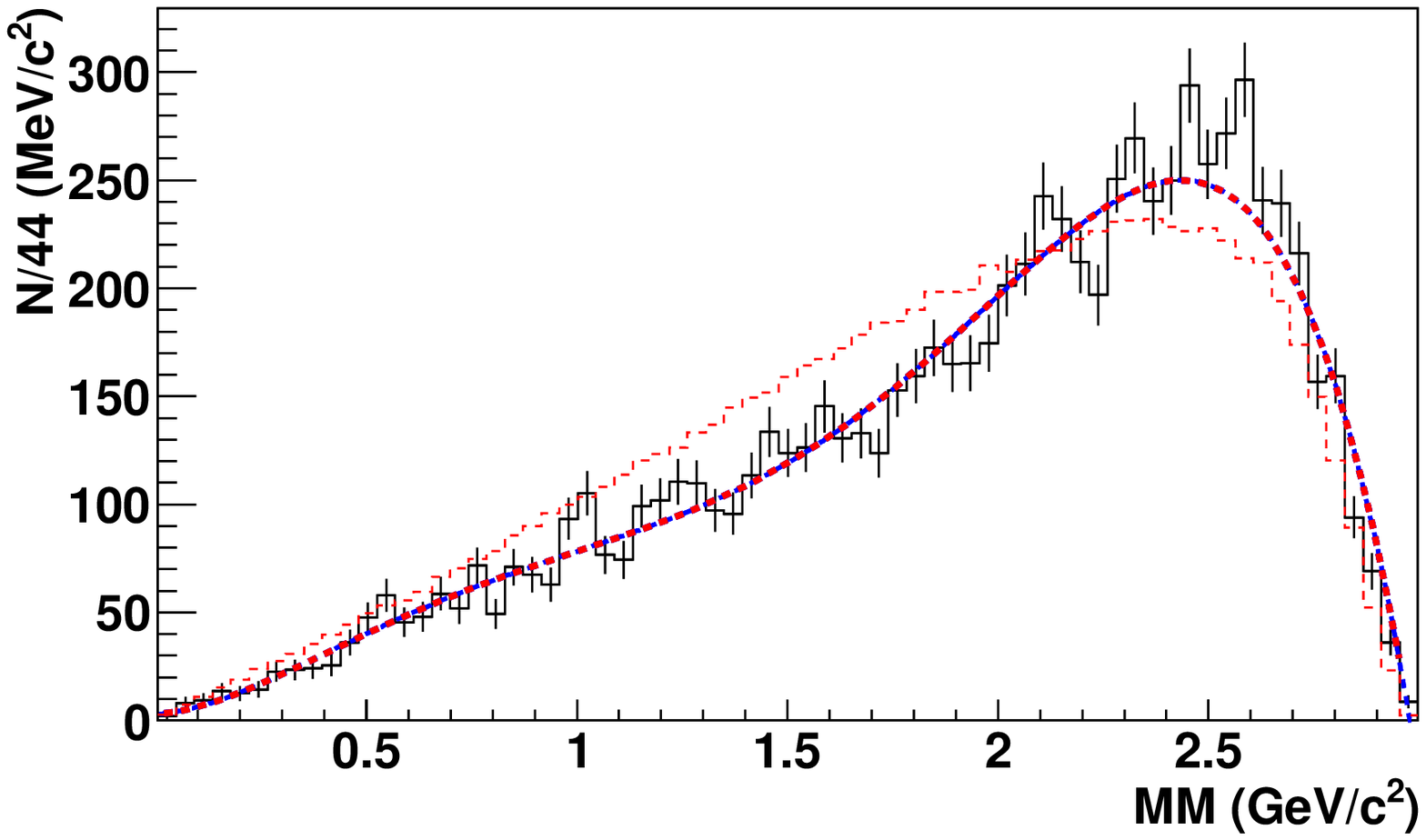}
     {\large b}
     \includegraphics[width=0.25\textwidth,height=0.15\textheight]{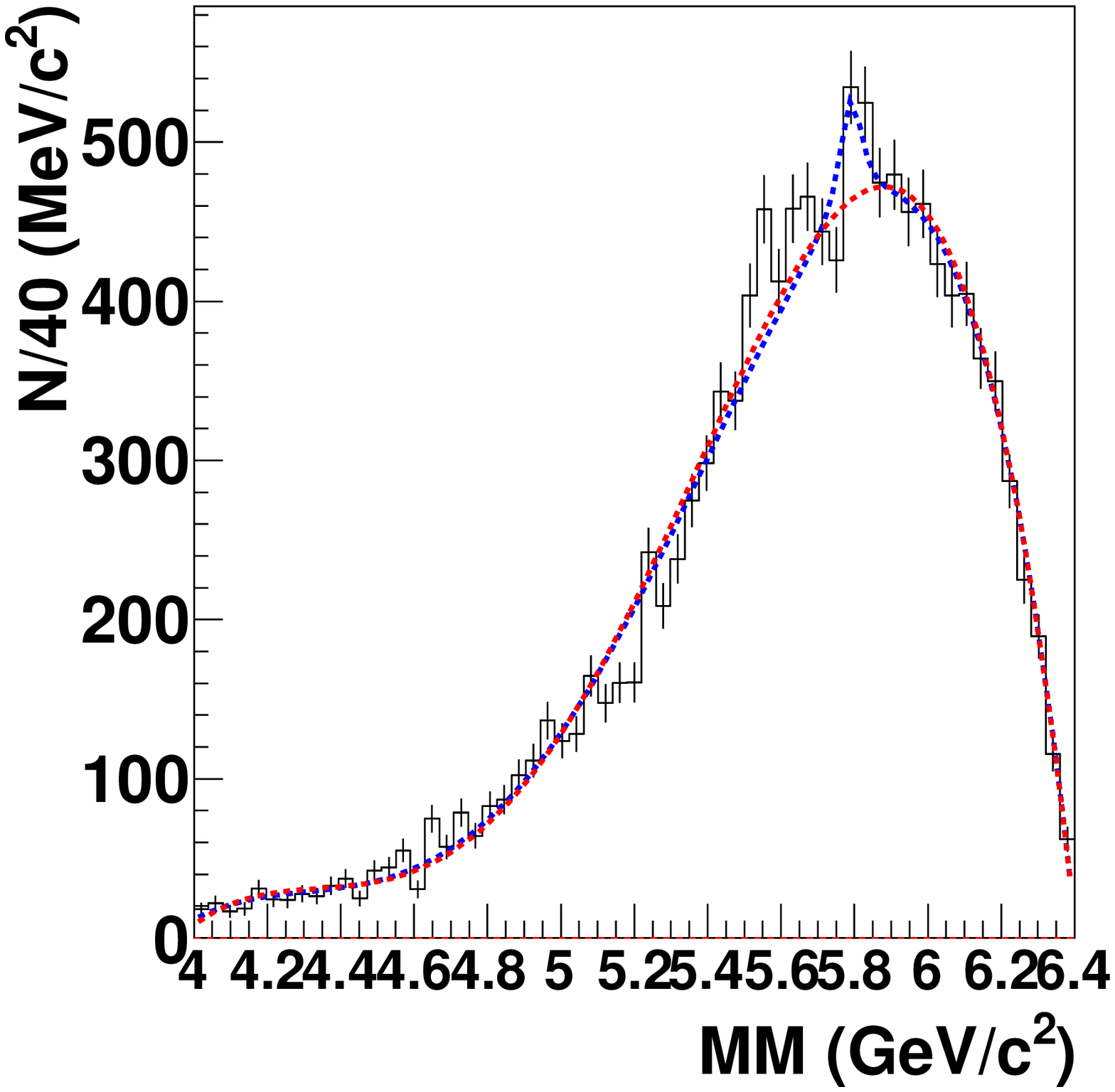}
     {\large c}
\caption{a)The missing mass spectrum(MM) of the (p,$K^0_s \pi^-$) reaction. b)The MM spectrum for the p(p,$K^0_s p$) reaction (7150 events). The dashed histogram is simulation by FRITIOF model. c)The MM spectrum  for the $^3H$(p,$K^0_s$ p) reaction(11118 events).}
\label{mm}       
\end{figure}

Fig. \ref{mm},a shows the missing mass(MM) spectrum for 3428 events  in the p(p,$K^0_s \pi^-$) reaction with a bin size of 34  MeV/$c^2$.
The curve(Fig.~\ref{mm},a is the sum of the background by the 9 order polynomial and 1 Breit-Wigner function. The peaks in the MM range of 3.35 GeV/$c^2$ with $\Gamma_{exp}\approx$ 90 MeV/$c^2$, S.D.( statistical  deviation)$5.7\sigma$($\approx$ 90 events in peak) and $\approx$ 3.00 GeV/$c^2$ with S.D.= 4.5$\sigma$ ($\approx$40-50 events in peak.) have been observed.

Fig.~\ref{mm},b  shows the MM spectrum for p(p,$K^0_s p$) reaction (7150 events) with a bin size of 44 MeV/$c^2$. There are observed signals
in the MM range of 1020, 2050 and 2580 MeV/$c^2$, with S.D.$\approx$ 4$\sigma$\cite{panda11}. A signal is not observed in the MM spectrum for the p(p,$K^0_s p$) reaction by FRITIOF model. Fig. \ref{mm},c shows the MM spectrum for 11118 events  in the $^3H$(p,$K^0_s$ p) reaction with a bin size of 40  MeV/$c^2$. The curve(Fig.~\ref{mm},c) is the sum of the background by the 9 order polynomial and 1 Breit-Wigner function. There is observed
signal in mass range of $^6_{\Lambda}He$(5.78 GeV/$c^2$) with $\Gamma$=90 MeV/$c^2$,  S.D.=5.7$\sigma$(120-150 events in the peak).
Fig. \ref{mml},a  shows the MM spectrum for 4148 events  in the $^3H$(p,$K^0_s$ p)(with $\pi^-$ events) reaction with a bin size of 40 MeV/$c^2$.
The same peak in the mass range of 5.8 GeV/$c^2$ ($^6_{\Lambda}He$) have been observed. There is also the significant enhancement in the MM range of 5.6 GeV/$c^2$ (interpreted as $^6He$).

 \section{The  missing mass spectra with $\Lambda$ hyperon}
\label{sec:2}

 Fig.~\ref{mml},b  shows  the  momentum distribution for $\pi^-$ with a bin size of 33 MeV/c for $^{12}C(p,\Lambda$)reaction. The fluctuation  is observed in the momentum range of 100-200 MeV/c. The 9-order polynomial did not describe the momentum distribution for $\pi^-$ in Fig.~\ref{mml},b.

Fig.~\ref{mml},c  shows  the  MM spectrum for the p(p,$\pi^-$) reaction (for events with $\Lambda$) with a bin size of 30 MeV/$c^2$. The background is the 9-order polynomial function. The peak in the MM range of 4200 MeV/$c^2$ with S.D.= 3.8$\sigma$ have been observed. There are observed small
 signals in the MM ranges of $^4_2He$(3.8 GeV/$c^2$) and $^4_{\Lambda}He$(3.9 GeV/$c^2$)(Fig.~\ref{mml},c).

 Fig.~\ref{mml},d  shows  the  MM spectrum for the p(p,$\Lambda$) reaction (with $\pi^- and \gamma$ events) with a bin size of 41 MeV/$c^2$. The background is  the 9-order polynomial function. The peak in MM range of 3200 MeV/$c^2$ with $\Gamma_{exp}\approx$90 MeV/$c^2$, S.D.= 6$\sigma$ have been observed. The same peak is observed from this data in\cite{pepan09} and OBELIX data in\cite{obelix}, what had been interpreted as S=-1 tribaryon states.

\section{Conclusion}
\label{sec:5}

\begin{figure*}
          \includegraphics[width=0.25\textwidth,height=0.15\textheight]{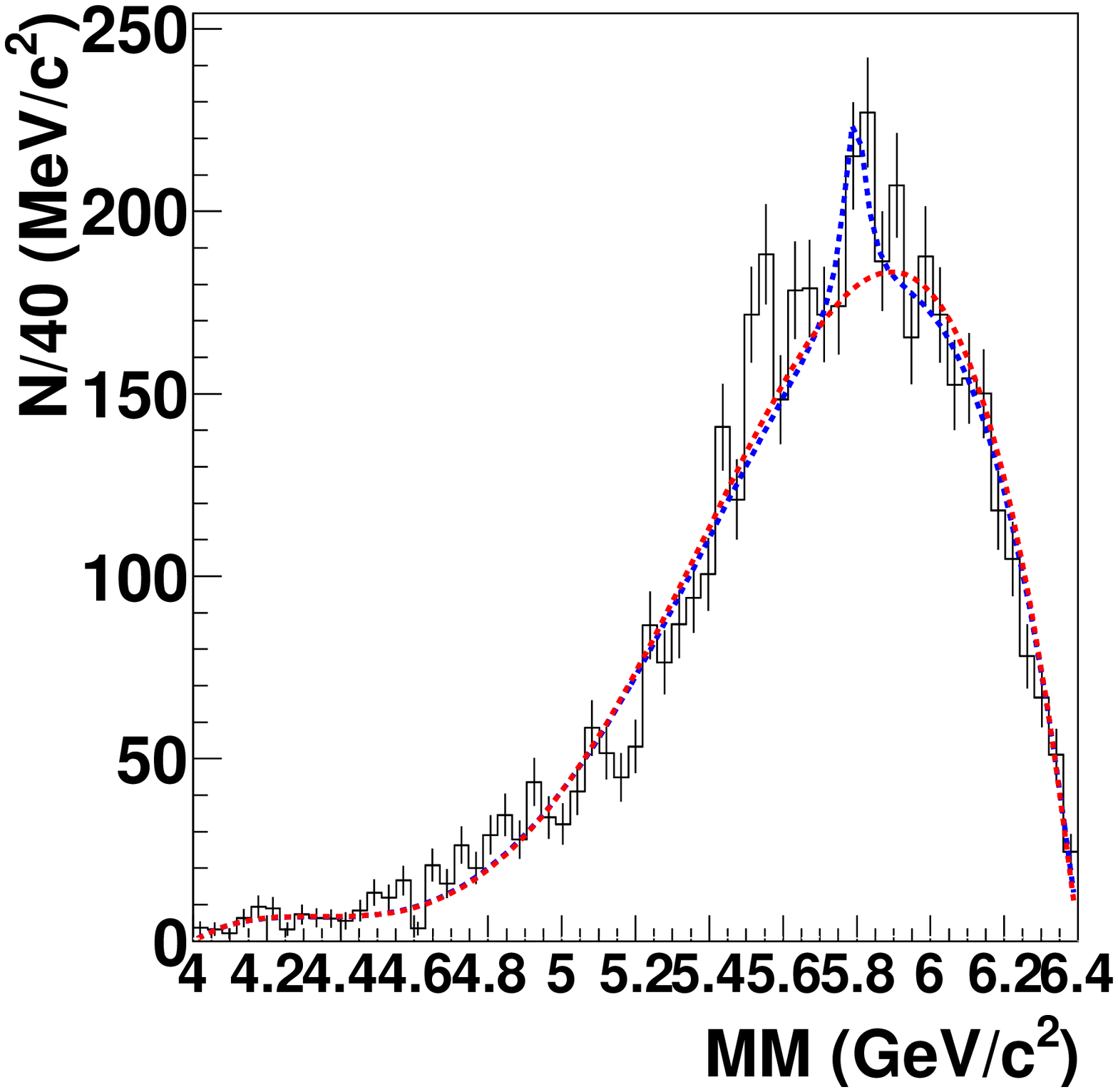}
     {\large a}
     \includegraphics[width=0.2\textwidth,height=0.15\textheight]{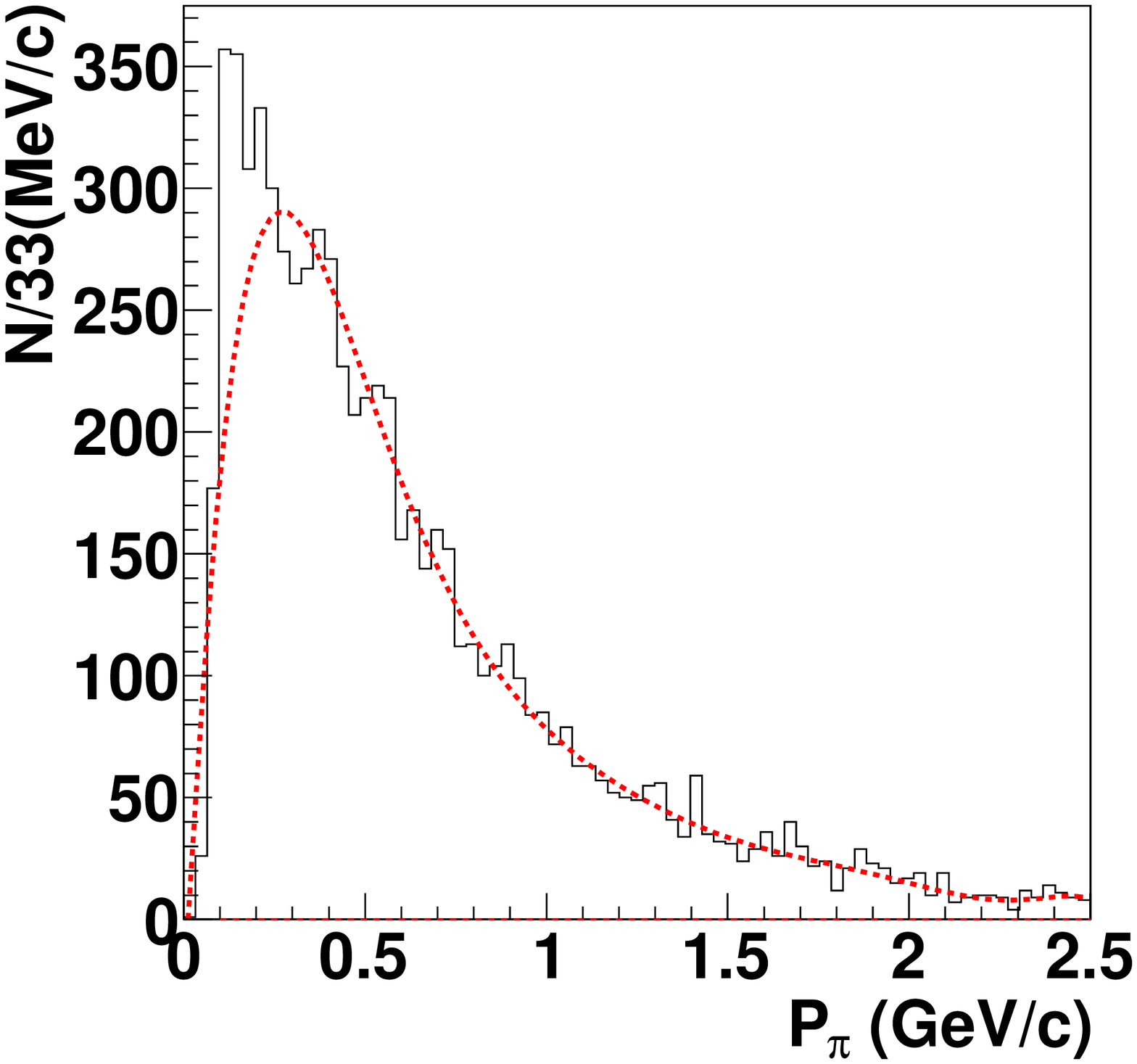}
     {\large b}
     \includegraphics[width=0.2\textwidth,height=0.15\textheight]{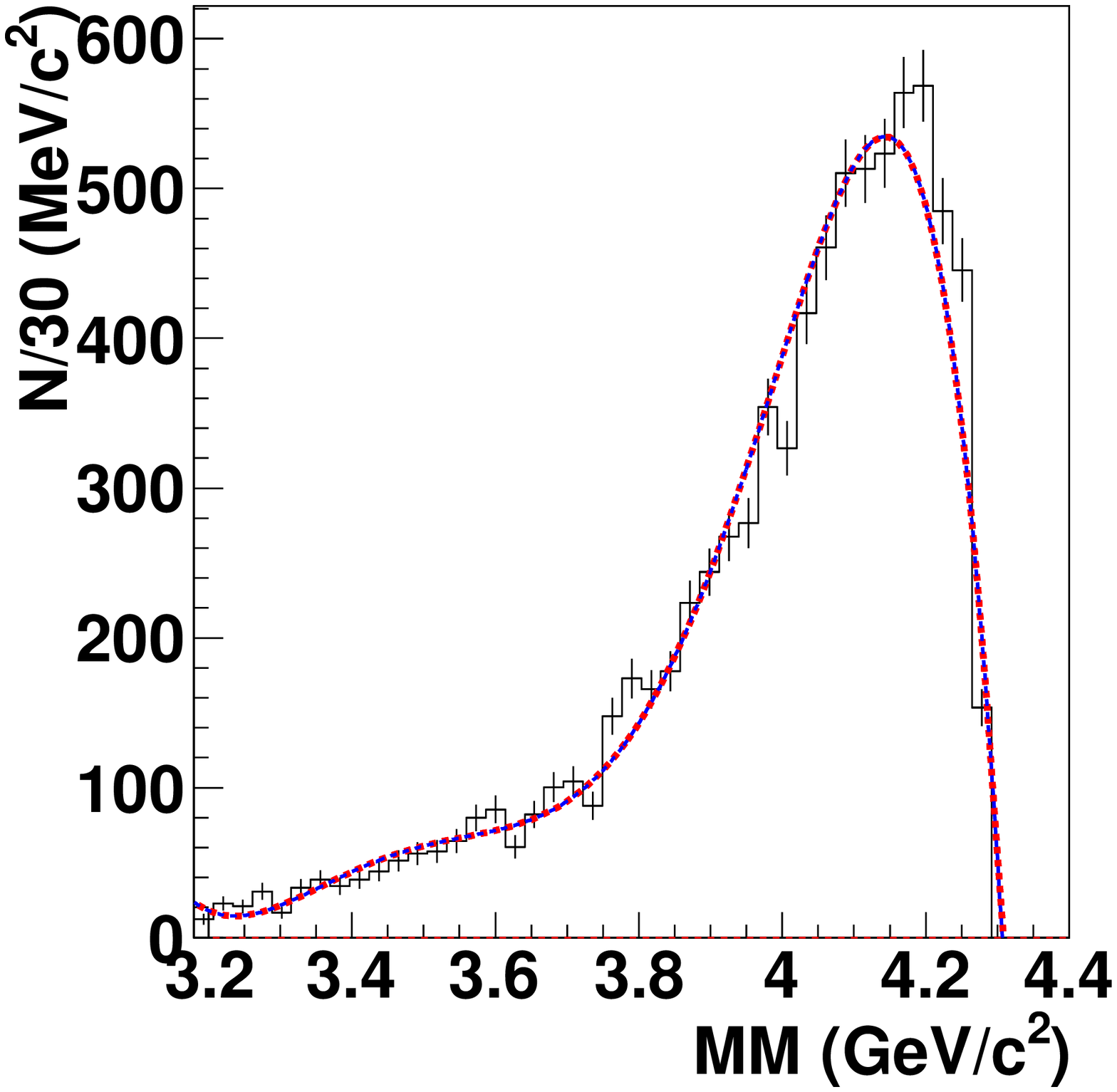}
     {\large c}
     \includegraphics[width=0.25\textwidth,height=0.15\textheight]{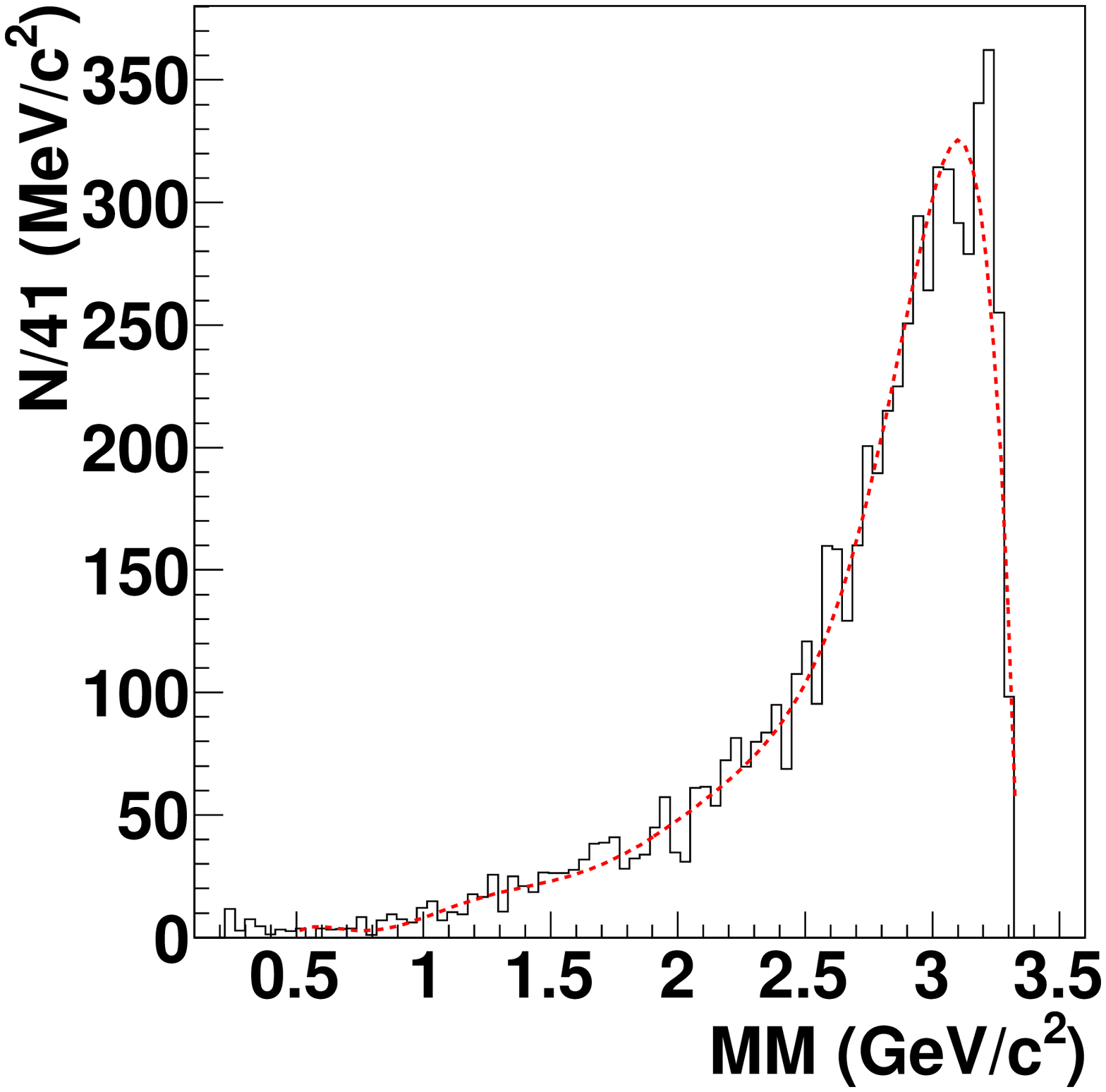}
     {\large d}
\caption{a)The MM spectrum for  the $^3H$(p,$K^0_s$ p) reaction (with $\pi^-$,4148 events).
b)The  momentum distribution for $\pi^-$ in the p+C$\to \Lambda \pi^-$ X reaction. c)The MM spectrum for  the p(p,$\pi^-$) reaction (with $\Lambda$, 7400 events). d)The MM spectrum for  the p(p,$\Lambda$) reaction (with $\pi^-, \gamma$, 5659 events).}
\label{mml}
\end{figure*}

The signals in the MM spectra have been observed only  a few possible channels from these  $^{12}C(p, K^0_s)$ and $^{12}C(p,\Lambda)$ reactions. These signals for the same reactions have observed with $\pi^-$ events too. The momentum spectrum of $\pi^-$ in range of 100-200 MeV/c from the   p+C$\to\pi^-\Lambda$X(p+C$\to\pi^-K^0_s$X) reactions have observed the significant enhancement(Fig.\ref{mml},b). Then the event-by-event analysis of $^{12}C(p, K^0_s)$ and $^{12}C(p,\Lambda)$ collisions  will the next step what allow identify a channel of reactions by kinematic fits for different hypothesis.

\end{document}